# Modernizing the SI – implications of recent progress with the fundamental constants

Nick Fletcher, Richard S. Davis, Michael Stock and Martin J.T. Milton,

Bureau International des Poids et Mesures (BIPM), Pavillon de Breteuil, 92312 Sèvres CEDEX, France.

e-mail : nick.fletcher@bipm.org

**Abstract**

Recent proposals to re-define some of the base units of the SI make use of definitions that refer to fixed numerical values of certain constants. We review these proposals in the context of the latest results of the least-squares adjustment of the fundamental constants and against the background of the difficulty experienced with communicating the changes. We show that the benefit of a definition of the kilogram made with respect to the atomic mass constant ($m_u$) may now be significantly stronger than when the choice was first considered 10 years ago.

## Introduction

The proposal to re-define four of the base units of the SI with respect to fixed numerical values of four constants has been the subject of much discussion and many publications. Although the possibility had been foreseen in publications during the 1990's [1] they were not articulated as a complete set of proposals until 2006 [2]. Subsequently, the General Conference on Weights and Measures, the forum for decision making between the Member States of the BIPM on all matters of measurement science and measurement units, addressed the matter at its 23rd meeting in 2007. It recognised the importance of considering such a re-definition, and invited the NMIs to "come to a view on whether it is possible". More recently at its 24th meeting in 2011, it noted that progress had been made towards such a re-definition and invited a final proposal when the experimental data were sufficiently robust to support one. The resolution also invited

> *'the CIPM to continue its work towards improved formulations for the definitions of the SI base units in terms of fundamental constants, having as far as possible a more easily understandable description for users in general, consistent with scientific rigour and clarity'.*

In response to this resolution, we have reviewed some of the possible formulations, for a re-definition of four base units (the kilogram, ampere, mole and kelvin). These have been considered previously in an unpublished paper discussed at the CCU in 2007 [3] and partially in a publication the same year [4].

This paper has been developed at a time when very good technical progress has been reported towards the adoption of new definitions for the base units of the SI [5]. In particular the latest least-squares adjustment of the fundamental constants indicates that the measured values of the Planck constant may have reached the necessary level of agreement. Other activities relating to the demonstration of the practical realization of the kilogram by the watt balance and the Si-XRCD methods are underway. However, the process of establishing a globally-agreed formulation and communicating it to users as requested by the CGPM in 2011 is proving to be difficult. For this reason it is timely to consider possible formulations for a re-definition in view of the recent results of the CODATA 2014 adjustment of the fundamental constants, which became available online in June 2015 [6]. (The full paper for the 2014 adjustment is not expected before the end of 2015, but a summary of recent advances is available [7]).

The previous analysis of possible formulations [3, 4] was based on the CODATA 2006 data set, and the 2005 proposals were prepared using the CODATA 2002 data. In addition to reduced uncertainties for most of the constants since those publications, the CODATA 2014 results illustrate a significantly improved level of confidence in the experimentally-determined value of the fine structure constant ($\alpha$) than was justified 10 years ago. These improvements, together with the recognition that abandoning the 1990 conventional values for of $K_J$ and $R_K$ will introduce unavoidable step changes in the realization of the electrical units [8] at the relative level of $-9.8\times10^{-8}$ and $+1.8 \times 10^{-8}$, provide a different context for the choice of formulation than was possible in 2007.

In this paper, we review five possible formulations, including the present version of the SI, in the light of the experimental progress in the determination of the values of the relevant fundamental constants. In the following section we provide a short review of relevant improvements in the CODATA 2014. We calculate the uncertainties of a list of constants subject to four different formulations of the SI base units and discuss their possible impact.

## The adjustment of the fundamental constants and the consideration of alternative unit systems

The CODATA adjustment of the fundamental constants has been published every four years since 1998 [9]. Each publication has brought together the best available experimental data, with the best available theory. Table 1 shows the improvements over the last five published adjustments for a selection of experimental constants of most relevance to the redefinition of the SI. The values given are relative standard uncertainties, and are plotted in Figure 1. All values are determined within the formulation of the present SI. (The symbols used in this table, and throughout the paper, are defined in Annex A).

We use the CODATA 2014 data set to analyse the impact of different formulations on a redefined SI [3,4]. The discussion of these systems benefits from the insight that any definition of a base unit can be considered to be a statement that a certain quantity is fixed. For example, the definitions of the kilogram, the ampere, the kelvin and the mole in the present SI can be considered to be statements that the mass of the International Prototype of the Kilogram ($m(\mathcal{K})$), the magnetic constant ($\mu_0$), and the triple point of water ($T_{tpw}$), and the molar mass of $^{12}$C ($M(^{12}\text{C})$) are fixed.

We consider the five formulations listed in Table 2, labelled Systems A to E. System A is the present SI and is shown as a reference for the comparison. Systems B to E have in common that the present definitions of the second and the meter are not changed. Fixed numerical values of $k_B$ and $N_A$ are used to replace the kelvin and mole definitions. They also all remove the artefact kilogram definition by giving a fixed numerical value to one other constant. They differ in the choice of the two constants used to define the mass and electrical units from the possible set $\{h, e, m_u, \mu_0\}$ [1] [2]. System B is the revision to the SI proposed by Mills *et al.* [2] and noted in Resolution 1 of the 24th CGPM which provides zero uncertainty in the two constants used for quantum electrical standards. System E provides greater clarity in the definition of the mass unit by defining it with respect to a fixed value of the atomic mass constant ($m_u$). Systems C and D provide mixtures of the benefits brought by B and E and are not considered further because they would provide no compelling advantages for those most concerned with either electrical or mass measurements.

**Uncertainties of a set of fundamental constants in the five formulations**

It is not necessary to perform a complete least-squares adjustment of the constants to illustrate the effects of different choices of constants to define the base units. A simplified approach can yield valid results [10]. The relative standard uncertainties of a set of fundamental constants in the systems A to E are shown in Table 3. We illustrate in detail how we derive the uncertainties on $R_K$, $K_J$, $h$ and $e$ within System E starting with the relationships:

$$R_K = \frac{h}{e^2} \tag{1}$$

$$K_J = \frac{2e}{h} \tag{2}$$

$$\alpha = \frac{\mu_0 e^2 c}{2h} \tag{3}$$

$$m_u = \frac{2hR_\infty}{\alpha^2 A_r(e) c} \tag{4}$$

The quantities $\alpha$ and $A_r(e)$ are dimensionless, and thus their uncertainties must be independent of the choice of unit system. The Rydberg constant ($R_\infty$) is derived from spectroscopic measurements that depend only on the units metre and second, and are unaffected by the choices we are considering here. To determine the uncertainties of the 'non-fixed' constants in System E, we rearrange (3) and (4) to give expressions for $h$ and $e$ in terms of the fixed constants $m_u$ and $\mu_0$, and the three measured constants $\alpha$ and $A_r(e)$ and $R_\infty$.

$$h = \frac{\alpha^2 m_u A_r(e) c}{2R_\infty} \tag{5}$$

[1] From the six possible combinations, we examine only four. The pair $h$, $m_u$ is excluded because fixing the numerical values of both would redefine the second via the frequency $m_u c^2/h$. The pair $e$, $\mu_0$ does generate a complete and valid unit system, but would add no benefits over the four examined here.

[2] We do not consider the alternative of fixing $m_e$ rather than $m_u$ since it brings no useful simplification to the definition of the mole and further removes the kilogram definition from one based on the mass of a single atom of $^{12}C$.

$$e = \sqrt{\frac{\alpha^3 m_u A_r(\mathrm{e})}{\mu_0 R_\infty}} \tag{6}$$

Then we can calculate the two electrical constants as

$$R_K = \frac{\mu_0 c}{2\alpha} \tag{7}$$

$$K_J = \frac{2}{\sqrt{hR_K}} = 4\left(\frac{\alpha m_u A_r(\mathrm{e}) c^2 \mu_0}{R_\infty}\right)^{-1/2} \tag{8}$$

The relative uncertainties of System E can be calculated from equations (5) to (8). The uncertainties within the Systems B, C and D can be derived in a similar way.

**Review of the advantages of System E**

We note the distribution of uncertainties associated with System E, which gives a fixed numerical value to $m_u$ in combination with maintaining the present numerical value of $\mu_0$. There are clear advantages to this choice for the definitions of the kilogram and mole, but it was previously considered incompatible with the universal use of quantum Hall and Josephson standards in electrical metrology because the constants $R_K$ and $K_J$ do not have defined numerical values in this system. In the discussion below, we consider why this constraint might now be reconsidered, and review the possible benefits of System E.

*The needs of mass metrology*

A detailed analysis of the options for the definition of the kilogram has been published [4, 11], but only in combination with a fixed value of $e$. In fact, the choice between $h$ or $m_u$ as a defining constant has no impact on the uncertainties achievable in the dissemination of macroscopic mass. This is because the ratio $h/m_u$ is now known with relative uncertainty $4.5\times10^{-10}$, which is negligible compared to the uncertainties achievable by either the watt balance [5] or silicon XRCD [5, 11] routes to the realization of the kilogram at a macroscopic scale. This was not true based on the 2002 data set, when many of the defining concepts for a possible SI redefinition were established. Two distinct arguments were made to prefer $h$ over $N_A$ for the kilogram redefinition[3] (as first expressed in [1]), that have been superseded by recent experimental progress. Firstly, the diversity and accuracy of silicon XRCD experiments has significantly improved and the availability of highly-pure spheres of isotopically enriched silicon will increase in the near future [12]. This route to the realization of the kilogram after redefinition will be at least as important as watt balances. Secondly, the improved uncertainty of $h/m_u$ [7, 13] means that both techniques can realize the kilogram equally well regardless of the choice made for the definition.

[3] The early discussion of a formulation of the definition of the kilogram with respect to a fixed value of $N_A$ corresponds to either System C or E discussed here.

There is, however, a strong argument in favour of fixing $m_u$. It allows a simple, readily understandable definition of the unit of mass expressed in terms of a constant which is itself a mass. The general public is familiar with the atomic theory of matter, and can visualize a kilogram defined by a specified number of identical atoms. By contrast, when $h$ is introduced as a 'defining constant', there is no straightforward way to present the mass unit; quantum physics must be invoked at some point, and this puts the base unit definition beyond the scope of accessible explanations. This disadvantage can certainly be justified if the more complex definition is necessary to obtain significant benefits for precision measurement. We suggest that the balance of this consideration should be reviewed given our updated analysis of the needs of electrical metrology presented below.

We note that the choice of defining constant has little or no impact on the technical preparations for the redefinition of the kilogram. The CCM/CCU roadmap for implementation would remain unchanged, the CCM conditions [11] for acceptance are unaffected, and the *mise en pratique* could readily be adapted. The only effect is in the communication of the change, where the simplified definition would be a significant advantage.

*The needs of chemical metrology*

An aspect of the possible adoption of System B that has created unexpected objections is the proposal to re-define the mole based on a fixed value of $N_A$ hence defining it as a fixed number of entities rather than as a mass of material equal to the atomic weight. The concerns have centred on the diminished importance of the molar mass constant ($M_u = M(^{12}C)/12$) resulting from it becoming an experimentally determined quantity; an objection that could be expressed as a concern about a lack of coherence in the system. Although there are no practical implications for measurements involving amount of substance or molar quantities, the introduction of a non-exact molar mass constant $M_u$ is not easy to accommodate in teaching. The choice of $m_u$ as the constant that defines the mass unit removes this difficulty and immediately allows the molar mass and the Avogadro constants to have fixed values; one may maintain both the present relation $M_u = 1\ \mathrm{g\ mol}^{-1}$ and have a defined numerical value for the Avogadro constant, $N_A$, which satisfies the relation $N_A = M_u/m_u$. This combination of fixed constants directly addresses the concerns raised about the mole being defined according to System B.

*The needs of electrical metrology*

A re-definition of the SI base units based on System B solves the present problem of electrical metrology, namely the existence of conventional 1990 units, which are no longer coherent with their SI counterparts even within the relatively large uncertainties with which the latter are realized. The conventional values of $R_K$ and $K_J$ were within the uncertainties of the corresponding SI values at the time they were agreed in 1990, but that is not the case today. The necessary step to resolve this is that the 1990 conventional values are abandoned, and quantum Hall resistance and Josephson voltage standards return to using the SI values of the constants $R_K$ and $K_J$ respectively. This can be achieved if $R_K$ and $K_J$ have defined numerical values according to System B but this condition, whilst obviously being sufficient, is not perhaps necessary. What is required is that the values of $R_K$ and $K_J$ are known with uncertainty low enough and with a demonstrated stability so that these values can be used without practical impact on electrical measurements for the foreseeable future. We discuss whether System E now fulfils this condition.

In System E, the relative uncertainty of $R_K$ is the same as that of $\alpha$ (now $2.3 \times 10^{-10}$) and that of $K_J$ is essentially equal to the uncertainty of $\sqrt{\alpha}$ ($1.2 \times 10^{-10}$). Our evidence base for suggesting that these values are adequate for the needs of electrical metrology comes from a study by a CCEM task group on the impact of moving from the 1990 values of $R_K$ and $K_J$ to the SI values. The required offsets are rather significant; using the CODATA 2014 values, the relative change in $R_K$ is $+1.8 \times 10^{-8}$; the relative change in $K_J$ is $-9.8\times10^{-8}$. These predicted changes were communicated to the NMIs and industrial users of electrical measurements at the CPEM and NCSLI conferences in 2014 [8]. The conclusion was that whilst these changes would be noticeable in some top-level laboratories, they could be accommodated with minimal practical disruption. The residual uncertainties allocated to $R_K$ and $K_J$ in System E are, respectively 100 and 1000 times smaller than these changes. We therefore argue that the usual metrological requirement of having a "safe" margin between the definitions of primary standards and the needs of the most demanding users is fulfilled.

Whilst having fixed numerical values for $R_K$ and $K_J$ would clearly be the first preference for electrical metrology, we argue that the residual uncertainties of System E are in fact small enough to make the distinction irrelevant. Relative uncertainties of $1.2\times10^{-10}$ and $2.3\times10^{-10}$ for $K_J$ and $R_K$ respectively can be included by default in the uncertainty budgets of even the best primary standards without consequence. (We contrast this to the presently applicable uncertainties linking the 1990 values and the SI, $4\times10^{-7}$ and $1\times10^{-7}$ respectively, which must be omitted by default in any precision electrical work.) The only situations where the System E uncertainties would be discernible are in direct comparisons of Josephson arrays or quantum Hall devices. In this case, the uncertainty would be correlated between the two systems and would not enter the final result of the comparison. We also note that whilst such tests of the underlying theory at the highest possible accuracy are of great interest, they are independent of the definitions of the unit system. They are also not representative of real world measurement needs.

The fine structure constant $\alpha$ plays a central role in determining the uncertainties claimed for $R_K$ and $K_J$, so it is reasonable to ask how reliable we believe the value of $\alpha$ to be. In System E future shifts in the CODATA value of $\alpha$ larger than its claimed uncertainty would result in inconvenient changes to the reference values of $R_K$ and $K_J$ used in electrical metrology. Between CODATA 2006 and 2010, the reported value for $\alpha$ did indeed change by 6.5 times the 2006 uncertainty (a relative shift of $4.4 \times 10^{-9}$). Fortunately, we can make the case that the reliability of experimental determinations of $\alpha$ has now greatly improved. The reason for the 2006 error is well understood (an error in the theoretical value of the electron magnetic–moment anomaly, $a_e$, which is no longer reliant on a single calculation). Importantly the value of $\alpha$ derived using $a_e$ has also since been confirmed by the completely independent, non QED route of atomic recoil experiments (*e.g.* measurements of $h/m(^{87}\mathrm{Rb})$) [9, 13]. The reliability of the measured value of $\alpha$ is key to the acceptability of System E; we can now have good reason to be confident in the CODATA 2014 uncertainty. We note also that we can expect improvements in $\alpha$ determinations to continue; this is a central and active area of both experimental and theoretical physics. After redefinition, System E would continue to benefit from these improvements with even smaller uncertainties on $K_J$ and $R_K$.

*Retaining the present exact value of the magnetic constant* $\mu_0$ ( $= 4\pi \times 10^{-7}$ N/A$^2$)

Discussion of System B has concluded that there are no practical problems with $\mu_0$ having an experimental value [2], especially as the uncertainty, being equal to that of $\alpha$, is negligible for any known application. We do not argue otherwise, but nonetheless draw attention to the advantages of maintaining the present fixed value. In the proposed SI based on System B, we would need to allow the value of the magnetic constant to change depending on experimental values and uncertainties. We parameterize this extra dependence of $\mu_0$ as $4\pi \times (1+\delta) \times 10^{-7}$ N/A$^2$, where the value of $\delta = (\alpha/\alpha_{2018} - 1)$, $\alpha_{2018}$ being the best available experimental value of $\alpha$ at the time of redefinition. The uncertainty of $\delta$ is the same as the relative uncertainty of $\alpha$, taking the value of $\alpha_{2018}$ to be exact. Thus if $\alpha = \alpha_{2018}$, $\delta = 0$ but $u(\delta) = u(\alpha)/\alpha_{2018}$. The difficulty is that whilst $\mu_0$ needs to be introduced in introductory texts in electromagnetism, the factor $(1+\delta)$ is difficult to explain without an advanced discussion of unit systems. This is similar to the problem posed by $M_u$ in texts treating molar quantities.

It has been noted that the deviations of $\mu_0$ and $M_u$ from the exact values $4\pi\times10^{-7}$ N/A$^2$ and 1 g mol$^{-1}$ will be small enough to be "ignored in practice" [2]. However, the familiar exact values are no longer strictly valid within an SI based on System B, and the explanation of why this is so involves a digression into topics beyond any introductory treatment, where both these constants are needed. This does not seem to be compatible with the desire of the CGPM for a system of units that is '*easily understandable for users in general, consistent with scientific rigour and clarity*'. System E maps these problems onto the numerical values for $R_K$ and $K_J$, which are only visible to readers of the *mise en pratique* for electrical units – an audience that is better equipped to understand the subtle compromises required in maintaining a coherent system of units.

We note also the relation between the SI and Gaussian or Lorentz-Heaviside unit systems. These cgs systems remain in widespread use [14]. In addition, the cgs-emu system was important when choosing the value and the unit of $\mu_0$ in the MKSA system, which later became the SI. In such systems that have no base dimension of an electrical nature, the fine structure constant is given by $\alpha = e^2/\hbar c$ . This is difficult to map on to a system that simultaneously defines numerical values for $c$, $e$ (in coulombs) and $h$. The factors used for conversion between unit systems [14] would have to be changed to avoid errors in high-precision calculations carried out in the cgs unit system. An example would be the conversion factor needed for the unit of charge (the statcoulomb) in the fine-structure constant when it is written in Gaussian units.

Although, again, there are no practical problems foreseen, an experimental value of $\mu_0$ risks pushing the SI further away from a unit system that can be readily reconciled with theoretical physics.

## Summary and conclusions

We have shown the impact of the CODATA 2014 adjustment of the fundamental constants on four possible formulations for a re-definition of the base units of the SI. We note that the arguments for and against the different formulations have changed since they were first considered in 2005 and subsequently proposed in 2006.

In particular, the advantages of System E as presented here are not new and do not individually make the case for a reconsideration of the plans for the adoption of a "new SI" based on System B. However, our analysis of the possible impact on electrical metrology presents a new perspective and we have shown that System E offers some clear advantages and overcomes some disadvantages that are intrinsic to System B. The advantages of System E are:

- It uses a definition of the kilogram that refers to a mass (in this case $m_u$). It is therefore easier to explain than one that refers to a fixed value of $h$ and does not necessitate recourse to an "explicit constant" formulation of the definitions, with its associated complexity. There will be no impact on the uncertainty with which the kilogram can be realised by the watt balance or silicon XRCD methods.
- It will be possible to define the mole with respect to a fixed value of the Avogadro constant, whilst also fixing the atomic mass constant ($m_u$) which is itself in universal use as the basis of the scale for atomic masses. Consequently, the molar mass constant ($M_u$) will continue to be fixed as it is at present. This is an approach that has been advocated for many years by practitioners in the chemical measurement community.
- Electrical measurements will be brought fully back within the SI and the magnetic constant ($\mu_0$) will continue to be fixed. The straightforward conversion between the SI formulation of the electrical units and those in the Gaussian system will remain. The uncertainty in the experimentally determined values in $K_J$ and $R_K$ will not place any limit on the practical use of Josephson voltage standards and quantum Hall resistance standards for high-accuracy electrical metrology. Future improvements in the experimental determination of the fine-structure constant, which can be expected to result from innovative new experimental methods, will in turn feed through to reduced uncertainties.

In summary, the choice of System E would bring the process of the search for an improved formulation for the definition of the kilogram and the mole back to the idea published in 1974 [15] and formulated into two definitions in 1999 [1] that:

> The kilogram is the mass of 5.018 450 XX $\times 10^{25}$ free $^{12}$C atoms at rest and in their ground state.
>
> The mole is the amount of substance of a system that contains 6.022 140 XX $\times 10^{23}$ specified entities.

In this formulation, the definition of the ampere can remain unchanged whilst allowing realisation of the electrical units through the electrical quantum metrology effects with a practical level of uncertainty unaffected by the residual uncertainties of $K_J$ and $R_K$. We believe that System E will provide an experimentally based, accessible and comprehensible system that is also coherent and scientifically correct [16].

## Annex A: Symbols and quantities

| | |
|---|---|
| $\alpha$ | fine structure constant ( $= \mu_0 e^2 c/2h$) |
| $K_J$ | Josephson constant ( $= 2e/h$) |
| $R_K$ | Von Klitzing constant ( $= h/e^2$) |
| $h$ | Planck constant |
| $e$ | elementary charge |
| $c$ | speed of light in vacuum |
| $\mu_0$ | magnetic constant (also commonly known as permeability of free space) |
| $m(K)$ | mass of the international prototype kilogram |
| $m_u$ | atomic mass constant ( $= m(^{12}C)/12$) |
| $M_u$ | molar mass constant |
| $m_e$ | mass of the electron |
| $A_r(e)$ | the relative atomic mass of the electron ( $= m_e/m_u$) |
| $R_\infty$ | Rydberg constant ( $= \alpha^2 m_e c/2h$) |
| $k_B$ | Boltzmann constant |
| $T_{tpw}$ | temperature of the triple point of water |
| $N_A$ | Avogadro constant |

**Table 1**: Relative standard uncertainties for selected constants from successive CODATA adjustments, as calculated within the present SI. The data are shown in graphical form in Figure 1.

| Date | $k_B$ | $h$ | $e$ | $N_A$ | $h/m_u$ | $\alpha$ | $A_r(e)$ | $R_\infty$ | $R_K$ | $K_J$ |
|---|---|---|---|---|---|---|---|---|---|---|
| **1998** | $1.7\times10^{-6}$ | $7.8\times10^{-8}$ | $3.9\times10^{-8}$ | $7.8\times10^{-8}$ | $7.6\times10^{-9}$ | $3.7\times10^{-9}$ | $2.1\times10^{-9}$ | $7.6\times10^{-12}$ | $3.7\times10^{-9}$ | $3.9\times10^{-8}$ |
| **2002** | $1.8\times10^{-6}$ | $1.7\times10^{-7}$ | $8.5\times10^{-8}$ | $1.7\times10^{-7}$ | $6.7\times10^{-9}$ | $3.3\times10^{-9}$ | $4.4\times10^{-10}$ | $6.6\times10^{-12}$ | $3.3\times10^{-9}$ | $8.5\times10^{-8}$ |
| **2006** | $1.7\times10^{-6}$ | $5.0\times10^{-8}$ | $2.5\times10^{-8}$ | $5.0\times10^{-8}$ | $1.4\times10^{-9}$ | $6.8\times10^{-10}$ | $4.2\times10^{-10}$ | $6.6\times10^{-12}$ | $6.8\times10^{-10}$ | $2.5\times10^{-8}$ |
| **2010** | $9.1\times10^{-7}$ | $4.4\times10^{-8}$ | $2.2\times10^{-8}$ | $4.4\times10^{-8}$ | $7.0\times10^{-10}$ | $3.2\times10^{-10}$ | $4.0\times10^{-10}$ | $5.0\times10^{-12}$ | $3.2\times10^{-10}$ | $2.2\times10^{-8}$ |
| **2014** | $5.7\times10^{-7}$ | $1.2\times10^{-8}$ | $6.1\times10^{-9}$ | $1.2\times10^{-8}$ | $4.5\times10^{-10}$ | $2.3\times10^{-10}$ | $2.9\times10^{-11}$ | $5.9\times10^{-12}$ | $2.3\times10^{-10}$ | $6.1\times10^{-9}$ |

**Table 2**: The definition of the five systems considered in this paper. System A corresponds to the present SI.

| System | Quantities with exact numerical values |
|---|---|
| A | $m(\mathcal{K})$, $\mu_0$, $T_{tpw}$, and $M(^{12}C)$ |
| B | $h$ , $e$ , $k_B$ and $N_A$. |
| C | $m_u$, $e$ , $k_B$ and $N_A$ |
| D | $h$ , $\mu_0$ , $k_B$ and $N_A$ |
| E | $m_u$ , $\mu_0$ , $k_B$ and $N_A$ |

**Table 3**: Relative standards uncertainties calculated for a selection of constants subject to the differing constraints of the five systems considered here. Calculations use data from CODATA-2014 [6]. System A corresponds to the present SI.

| Quantity | System A | System B | System C | System D | System E |
|---|---|---|---|---|---|
| $m(\mathcal{K})$ | 0 | $1.2\times10^{-8}$ | $1.2\times10^{-8}$ | $1.2\times10^{-8}$ | $1.2\times10^{-8}$ |
| $h$ | $1.2\times10^{-8}$ | 0 | $4.5\times10^{-10}$ | 0 | $4.5\times10^{-10}$ |
| $e$ | $6.1\times10^{-9}$ | 0 | 0 | $1.2\times10^{-10}$ | $3.5\times10^{-10}$ |
| $\alpha$ | $2.3\times10^{-10}$ | $2.3\times10^{-10}$ | $2.3\times10^{-10}$ | $2.3\times10^{-10}$ | $2.3\times10^{-10}$ |
| $R_K$ | $2.3\times10^{-10}$ | 0 | $4.5\times10^{-10}$ | $2.3\times10^{-10}$ | $2.3\times10^{-10}$ |
| $K_J$ | $6.1\times10^{-9}$ | 0 | $4.5\times10^{-10}$ | $1.2\times10^{-10}$ | $1.2\times10^{-10}$ |
| $\mu_0$ | 0 | $2.3\times10^{-10}$ | $6.9\times10^{-10}$ | 0 | 0 |
| $m_u$ | $1.2\times10^{-8}$ | $4.5\times10^{-10}$ | 0 | $4.5\times10^{-10}$ | 0 |
| $m_e$ | $1.2\times10^{-8}$ | $4.5\times10^{-10}$ | $2.9\times10^{-11}$ | $4.5\times10^{-10}$ | $2.9\times10^{-11}$ |
| $N_A$ | $1.2\times10^{-8}$ | 0 | 0 | 0 | 0 |
| $M(^{12}C)$ | 0 | $4.5\times10^{-10}$ | 0 | $4.5\times10^{-10}$ | 0 |
| $k_B$ | $5.7\times10^{-7}$ | 0 | 0 | 0 | 0 |
| $T_{tpw}$ | 0 | $5.7\times10^{-7}$ | $5.7\times10^{-7}$ | $5.7\times10^{-7}$ | $5.7\times10^{-7}$ |

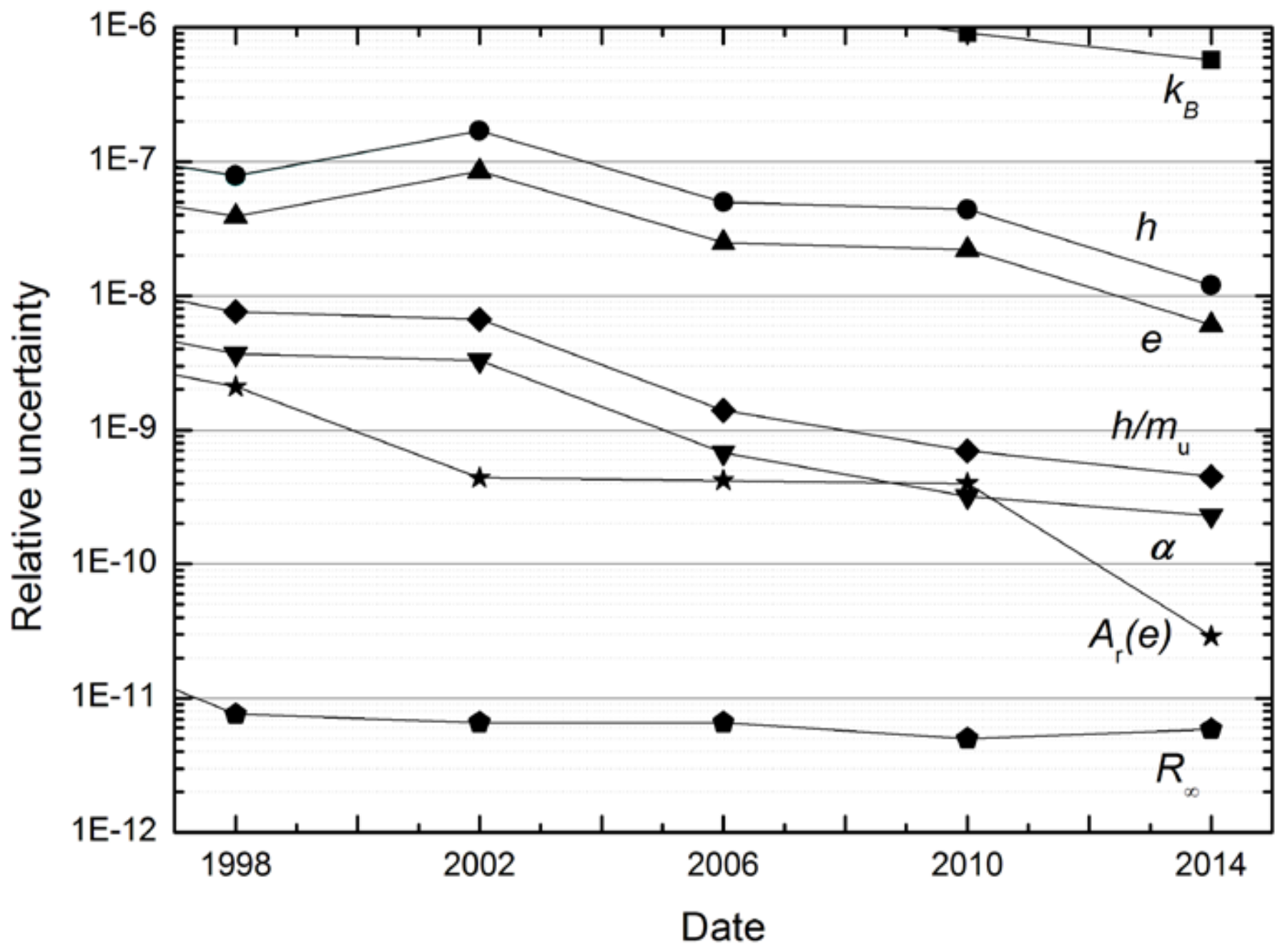

**Figure 1**: Evolution of the relative standard uncertainties of the constants shown in Table 1.